\documentclass[%
 twocolumn,
 aps,pra,
 amsmath,amssymb,
 superscriptaddress,
 nofootinbib,
 reprint,%
]{revtex4-1}
\usepackage{stmaryrd}
\usepackage{color}
\usepackage{array}
\usepackage{enumitem}
\usepackage{mathpazo}
\usepackage{setspace}
\usepackage{bm}
\usepackage{amssymb}
\usepackage{amsthm}
\usepackage{mathtools}
\usepackage{multirow}
\usepackage{cases}
\usepackage[colorlinks=true,linkcolor=blue,citecolor=blue,pdfauthor={ },pdftitle={ },pdfsubject={ },pdfkeywords={ }]{hyperref}
\usepackage{pgfplots}
\usepackage{lipsum}
\usepackage{youngtab}
\usepackage{booktabs}

\newtheorem{definition}{Definition}
\newtheorem{proposition}[definition]{Proposition}
\newtheorem{lemma}[definition]{Lemma}

\newtheorem{theorem}[definition]{Theorem}

\newtheorem{corollary}[definition]{Corollary}
\newtheorem{conjecture}[definition]{Conjecture}

\newtheorem{remark}[definition]{Remark}
\newtheorem{example}[definition]{Example}
\newtheorem{question}[definition]{Question}

\def\Dbar{\leavevmode\lower.6ex\hbox to 0pt
{\hskip-.23ex\accent"16\hss}D}
\makeatletter
\def\url@leostyle{%
  \@ifundefined{selectfont}{\def\UrlFont{\sf}}{\def\UrlFont{\small\ttfamily}}}
\makeatother
\DeclareMathOperator{\Tr}{Tr} %

\def\bcj{\begin{conjecture}}
\def\ecj{\end{conjecture}}
\def\bcr{\begin{corollary}}
\def\ecr{\end{corollary}}
\def\bd{\begin{definition}}
\def\ed{\end{definition}}
\def\bea{\begin{eqnarray}}
\def\eea{\end{eqnarray}}
\def\bem{\begin{enumerate}}
\def\eem{\end{enumerate}}
\def\bex{\begin{example}}
\def\eex{\end{example}}
\def\bim{\begin{itemize}}
\def\eim{\end{itemize}}
\def\bl{\begin{lemma}}
\def\el{\end{lemma}}
\def\bpf{\begin{proof}}
\def\epf{\end{proof}}
\def\bpp{\begin{proposition}}
\def\epp{\end{proposition}}
\def\bqu{\begin{question}}
\def\equ{\end{question}}
\def\br{\begin{remark}}
\def\er{\end{remark}}
\def\bt{\begin{theorem}}
\def\et{\end{theorem}}

\def\btb{\begin{tabular}}
\def\etb{\end{tabular}}

\newcommand{\nc}{\newcommand}

 \nc{\bA}{{\bf A}} \nc{\bB}{{\bf B}} \nc{\bC}{{\bf C}}
 \nc{\bD}{{\bf D}} \nc{\bE}{{\bf E}} \nc{\bF}{{\bf F}}
 \nc{\bG}{{\bf G}} \nc{\bH}{{\bf H}} \nc{\bI}{{\bf I}}
 \nc{\bJ}{{\bf J}} \nc{\bK}{{\bf K}} \nc{\bL}{{\bf L}}
 \nc{\bM}{{\bf M}} \nc{\bN}{{\bf N}} \nc{\bO}{{\bf O}}
 \nc{\bP}{{\bf P}} \nc{\bQ}{{\bf Q}} \nc{\bR}{{\bf R}}
 \nc{\bS}{{\bf S}} \nc{\bT}{{\bf T}} \nc{\bU}{{\bf U}}
 \nc{\bV}{{\bf V}} \nc{\bW}{{\bf W}} \nc{\bX}{{\bf X}}
 \nc{\bZ}{{\bf Z}}

\nc{\cA}{{\cal A}} \nc{\cB}{{\cal B}} \nc{\cC}{{\cal C}}
\nc{\cD}{{\cal D}} \nc{\cE}{{\cal E}} \nc{\cF}{{\cal F}}
\nc{\cG}{{\cal G}} \nc{\cH}{{\cal H}} \nc{\cI}{{\cal I}}
\nc{\cJ}{{\cal J}} \nc{\cK}{{\cal K}} \nc{\cL}{{\cal L}}
\nc{\cM}{{\cal M}} \nc{\cN}{{\cal N}} \nc{\cO}{{\cal O}}
\nc{\cP}{{\cal P}} \nc{\cQ}{{\cal Q}} \nc{\cR}{{\cal R}}
\nc{\cS}{{\cal S}} \nc{\cT}{{\cal T}} \nc{\cU}{{\cal U}}
\nc{\cV}{{\cal V}} \nc{\cW}{{\cal W}} \nc{\cX}{{\cal X}}
\nc{\cZ}{{\cal Z}}

\nc{\hA}{{\hat{A}}} \nc{\hB}{{\hat{B}}} \nc{\hC}{{\hat{C}}}
\nc{\hD}{{\hat{D}}} \nc{\hE}{{\hat{E}}} \nc{\hF}{{\hat{F}}}
\nc{\hG}{{\hat{G}}} \nc{\hH}{{\hat{H}}} \nc{\hI}{{\hat{I}}}
\nc{\hJ}{{\hat{J}}} \nc{\hK}{{\hat{K}}} \nc{\hL}{{\hat{L}}}
\nc{\hM}{{\hat{M}}} \nc{\hN}{{\hat{N}}} \nc{\hO}{{\hat{O}}}
\nc{\hP}{{\hat{P}}} \nc{\hR}{{\hat{R}}} \nc{\hS}{{\hat{S}}}
\nc{\hT}{{\hat{T}}} \nc{\hU}{{\hat{U}}} \nc{\hV}{{\hat{V}}}
\nc{\hW}{{\hat{W}}} \nc{\hX}{{\hat{X}}} \nc{\hZ}{{\hat{Z}}}

\newcommand{\bra}[1]{\langle#1|}
\newcommand{\ket}[1]{|#1\rangle}

\def\Dbar{\leavevmode\lower.6ex\hbox to 0pt
{\hskip-.23ex\accent"16\hss}D}
\begin{document}


\def\be{\begin{eqnarray}}
\def\ee{\end{eqnarray}}


\newcommand{\ca}{\mathcal A}

\newcommand{\cb}{\mathcal B}
\newcommand{\cc}{\mathcal C}
\newcommand{\cd}{\mathcal D}
\newcommand{\ce}{\mathcal E}
\newcommand{\cf}{\mathcal F}
\newcommand{\cg}{\mathcal G}
\newcommand{\ch}{\mathcal H}
\newcommand{\ci}{\mathcal I}
\newcommand{\cj}{\mathcal J}
\newcommand{\ck}{\mathcal K}
\newcommand{\cl}{\mathcal L}
\newcommand{\cm}{\mathcal M}
\newcommand{\cn}{\mathcal N}
\newcommand{\co}{\mathcal O}
\newcommand{\cp}{\mathcal P}
\newcommand{\cq}{\mathcal Q}
\newcommand{\calr}{\mathcal R}
\newcommand{\cs}{\mathcal S}
\newcommand{\ct}{\mathcal T}
\newcommand{\cu}{\mathcal U}
\newcommand{\cv}{\mathcal V}
\newcommand{\cw}{\mathcal W}
\newcommand{\cx}{\mathcal X}
\newcommand{\cy}{\mathcal Y}
\newcommand{\cz}{\mathcal Z}


\newcommand{\sa}{\mathscr{A}}
\newcommand{\sm}{\mathscr{M}}


\newcommand{\fa}{\mathfrak{a}}  \newcommand{\Fa}{\mathfrak{A}}
\newcommand{\fb}{\mathfrak{b}}  \newcommand{\Fb}{\mathfrak{B}}
\newcommand{\fc}{\mathfrak{c}}  \newcommand{\Fc}{\mathfrak{C}}
\newcommand{\fd}{\mathfrak{d}}  \newcommand{\Fd}{\mathfrak{D}}
\newcommand{\fe}{\mathfrak{e}}  \newcommand{\Fe}{\mathfrak{E}}
\newcommand{\ff}{\mathfrak{f}}  \newcommand{\Ff}{\mathfrak{F}}
\newcommand{\fg}{\mathfrak{g}}  \newcommand{\Fg}{\mathfrak{G}}
\newcommand{\fh}{\mathfrak{h}}  \newcommand{\Fh}{\mathfrak{H}}
\newcommand{\fraki}{\mathfrak{i}}       \newcommand{\Fraki}{\mathfrak{I}}
\newcommand{\fj}{\mathfrak{j}}  \newcommand{\Fj}{\mathfrak{J}}
\newcommand{\fk}{\mathfrak{k}}  \newcommand{\Fk}{\mathfrak{K}}
\newcommand{\fl}{\mathfrak{l}}  \newcommand{\Fl}{\mathfrak{L}}
\newcommand{\fm}{\mathfrak{m}}  \newcommand{\Fm}{\mathfrak{M}}
\newcommand{\fn}{\mathfrak{n}}  \newcommand{\Fn}{\mathfrak{N}}
\newcommand{\fo}{\mathfrak{o}}  \newcommand{\Fo}{\mathfrak{O}}
\newcommand{\fp}{\mathfrak{p}}  \newcommand{\Fp}{\mathfrak{P}}
\newcommand{\fq}{\mathfrak{q}}  \newcommand{\Fq}{\mathfrak{Q}}
\newcommand{\fr}{\mathfrak{r}}  \newcommand{\Fr}{\mathfrak{R}}
\newcommand{\fs}{\mathfrak{s}}  \newcommand{\Fs}{\mathfrak{S}}
\newcommand{\ft}{\mathfrak{t}}  \newcommand{\Ft}{\mathfrak{T}}
\newcommand{\fu}{\mathfrak{u}}  \newcommand{\Fu}{\mathfrak{U}}
\newcommand{\fv}{\mathfrak{v}}  \newcommand{\Fv}{\mathfrak{V}}
\newcommand{\fw}{\mathfrak{w}}  \newcommand{\Fw}{\mathfrak{W}}
\newcommand{\fx}{\mathfrak{x}}  \newcommand{\Fx}{\mathfrak{X}}
\newcommand{\fy}{\mathfrak{y}}  \newcommand{\Fy}{\mathfrak{Y}}
\newcommand{\fz}{\mathfrak{z}}  \newcommand{\Fz}{\mathfrak{Z}}

\newcommand{\cfg}{\dot \fg}
\newcommand{\cFg}{\dot \Fg}
\newcommand{\ccg}{\dot \cg}
\newcommand{\circj}{\dot {\mathbf J}}
\newcommand{\circs}{\circledS}
\newcommand{\jmot}{\mathbf J^{-1}}


\newcommand{\rmd}{\mathrm d}
\newcommand{\mca}{\ ^-\!\!\ca}
\newcommand{\pca}{\ ^+\!\!\ca}
\newcommand{\peq}{^\Psi\!\!\!\!\!=}
\newcommand{\lt}{\left}
\newcommand{\rt}{\right}
\newcommand{\HN}{\hat{H}(N)}
\newcommand{\HM}{\hat{H}(M)}
\newcommand{\Hv}{\hat{H}_v}
\newcommand{\cyl}{\mathbf{Cyl}}
\newcommand{\lag}{\left\langle}
\newcommand{\rag}{\right\rangle}
\newcommand{\Ad}{\mathrm{Ad}}
\newcommand{\trace}{\mathrm{tr}}
\newcommand{\bbc}{\mathbb{C}}
\newcommand{\AC}{\overline{\mathcal{A}}^{\mathbb{C}}}
\newcommand{\Ar}{\mathbf{Ar}}
\newcommand{\uc}{\mathrm{U(1)}^3}
\newcommand{\M}{\hat{\mathbf{M}}}
\newcommand{\spin}{\text{Spin(4)}}
\newcommand{\id}{\mathrm{id}}
\newcommand{\Pol}{\mathrm{Pol}}
\newcommand{\Fun}{\mathrm{Fun}}
\newcommand{\bp}{p}
\newcommand{\act}{\rhd}
\newcommand{\data}{\lt(j_{ab},A,\bar{A},\xi_{ab},z_{ab}\rt)}
\newcommand{\datao}{\lt(j^{(0)}_{ab},A^{(0)},\bar{A}^{(0)},\xi_{ab}^{(0)},z_{ab}^{(0)}\rt)}
\newcommand{\deltadata}{\lt(j'_{ab}, A',\bar{A}',\xi_{ab}',z_{ab}'\rt)}
\newcommand{\background}{\lt(j_{ab}^{(0)},g_a^{(0)},\xi_{ab}^{(0)},z_{ab}^{(0)}\rt)}
\newcommand{\sgn}{\mathrm{sgn}}
\newcommand{\vth}{\vartheta}
\newcommand{\rmi}{\mathrm{i}}
\newcommand{\bfmu}{\pmb{\mu}}
\newcommand{\bfnu}{\pmb{\nu}}
\newcommand{\bfm}{\mathbf{m}}
\newcommand{\bfn}{\mathbf{n}}


\newcommand{\sz}{\mathscr{Z}}
\newcommand{\sk}{\mathscr{K}}

\title{Tapping into Permutation Symmetry for Improved Detection of
$k$-Symmetric Extensions}

\author{Youning Li}%
\affiliation{College of Science, China Agricultural University, Beijing, 100080, People's Republic of China}

\author{Chao Zhang}%
\affiliation{Department of Physics, The Hong Kong University of
  Science and Technology, Clear Water Bay, Kowloon, Hong Kong, China}

\author{Shi-Yao Hou}

\affiliation{College of Physics and Electronic Engineering, Center for Computational Sciences,  Sichuan Normal University, Chengdu 610068, China}

\author{Zipeng Wu}%
\affiliation{Department of Physics, The Hong Kong University of
  Science and Technology, Clear Water Bay, Kowloon, Hong Kong, China}

\author{Xuanran Zhu}%
\affiliation{Department of Physics, The Hong Kong University of
  Science and Technology, Clear Water Bay, Kowloon, Hong Kong, China}

\author{Bei Zeng}
\affiliation{Department of Physics, The Hong Kong University of
  Science and Technology, Clear Water Bay, Kowloon, Hong Kong, China}%

\date{\today}

\begin{abstract}

Symmetric extensions are essential in quantum mechanics, providing a lens to investigate the correlations of entangled quantum systems and to address challenges like the quantum marginal problem. Though semi-definite programming (SDP) is a recognized method for handling symmetric extensions, it grapples with computational constraints, especially due to the large real parameters in generalized qudit systems. In this study, we introduce an approach that adeptly leverages permutation symmetry. By fine-tuning the SDP problem for detecting \( k \)-symmetric extensions, our method markedly diminishes the searching space dimensionality and trims the number of parameters essential for positive definiteness tests. This leads to an algorithmic enhancement, reducing the complexity from \( O(d^{2k}) \) to \( O(k^{d^2}) \) in the qudit \( k \)-symmetric extension scenario. Additionally, our approach streamlines the process of verifying the positive definiteness of the results. These advancements pave the way for deeper insights into quantum correlations, highlighting potential avenues for refined research and innovations in quantum information theory.
\end{abstract}

\maketitle
\renewcommand\theequation{\arabic{section}.\arabic{equation}}
\setcounter{tocdepth}{4}
\makeatletter
\@addtoreset{equation}{section}
\makeatother

\section{Introduction}


In the intricate domain of quantum mechanics, symmetric extensions stand out as a cornerstone, providing a structured mathematical lens to explore the nature and behavior of quantum states. A bipartite state \( \rho_{AB} \) is deemed symmetrically extendible if there exists a multi-partite density matrix \( \rho_{A_1A_2\ldots A_mB_1B_2\ldots B_n} \) such that each of its reduced density matrices, when traced over its complements, matches \( \rho_{AB} \):
\begin{equation}\label{Eq: constrain}
\Tr_{(A_jB_k)^c}(\rho_{A_1A_2\ldots A_mB_1B_2\ldots B_n})=\rho_{AB},~\forall j,k.
\end{equation}

Delving into the importance of symmetric extensions, they serve as a tangible framework to probe the nature of quantum entanglement, offering a means to understand the profound correlations present in entangled quantum systems~\cite{Horodecki2009,PhysRevLett.69.2881,NC00}. Furthermore, they pave the way for addressing the quantum marginal problem, which investigates the necessary conditions under which a set of density matrices can correspond to a global state~\cite{Kly06,chen2014symmetric}. This problem's universality is showcased by its resonance with the \( N \)-representability problem in quantum chemistry~\cite{Col63,Erd72}.

A common approach for identifying a $k$-extension is to cast the problem as a semi-definite programming (SDP) problem~\cite{navascues2009power,brandao2011quasipolynomial,johnston2014detection}. SDP, a form of convex optimization, involves minimizing a linear function subject to the constraints that the solution lies in the intersection of the cone of positive semidefinite matrices and an affine space. Given that density matrices are inherently semidefinite, SDP has found extensive application in quantum information problems ~\cite{vandenberghe1996semidefinite,doherty2004complete}.
By leveraging the properties of SDP, we can devise efficient algorithms for detecting $k$-extensions. For example, the algorithm used in QETLAB that determines if a bipartite quantum state  $\rho_{AB}$ is $k$-symmetric extendible for the $d$-dimensional subsystem $B$ has the form:
\begin{equation}\label{Eq: SDP_SE}
\begin{aligned}
    &\mathrm{find}\;\tilde{\rho}\\
    s.t.&\;\begin{cases}
        \tilde{\rho}\succeq0\\
        \mathrm{Tr}_{\left(AB_{1}\right)^{c}}\left(\tilde{\rho}\right)=\rho_{AB}\\
        \left(\mathbb{1}_{A}\otimes P_{ij}\right)\tilde{\rho}\left(\mathbb{1}_{A}\otimes P_{ij}\right)=\tilde{\rho}, \forall i,j
    \end{cases}
\end{aligned}
\end{equation}
where $\tilde{\rho}=\rho_{AB_1B_2\ldots B_k}$ is the $k$-symmetric extension of $\rho_{AB}$ and the operator $P_{ij}$ is an element in permutation group $S_k$. However, the substantial number of real parameters, notably in the general qudit scenario, can present formidable computational obstacles. This is largely due to the requirement of the entire extended Hilbert space $\mathcal{H}_{AB_1B_2\ldots B_k}$, which scales as $O((d^2)^{k})$. Such exponential scaling can make calculations intractable for larger systems or higher dimensions.

In this work we are going to present a new optimization scheme, which not only considers the permutation symmetry to reduce the total parameters but also optimize in the subroutine to determine positive definiteness, where the parameters for single time optimization grows no faster than
\[
\prod_{m=1}^{d-1}\frac{1}{m!}\left(1+\frac{2k}{d(d+1)}\right)^{d(d-1)/2},
\]
for large $k$ and $d$.

A testament to our methodology's effectiveness is its application to the renowned bipartite Werner state, where it exhibits a pronounced acceleration in comparison to the established QETLAB software. This enhancement equips us to approach larger \( k \)-extension challenges with unparalleled efficiency.

Furthermore, our calculations have explicitly determined the dimensions of the searching space and the number of parameters required for positive definiteness tests. This efficiency stems from our algorithm's ability to undergo multiple distinct positive-definiteness tests, each correlating to a unique Young diagram. Each individual test, though involving a significantly smaller matrix, culminates in a comprehensive and efficient analysis.

Our findings contribute to a clearer understanding of quantum systems, potentially aiding in the design of more proficient quantum algorithms and enhancing our grasp of quantum information theory.

The structure of our paper is as follows: Sec.~\ref{sec:2} delves into the intricacies of the 3-extension of the qutrit case as an illustrative example. Sec.~\ref{sec:3} elucidates our methodology to compute the reduced density matrix of global states for a general \( k \)-extendible state, and underscores our rationale for dimensionality reduction. Sec.\ref{sec:4} shows the comparison of our new algorithm and traditional one. Concluding insights and discussions are furnished in Sec.~\ref{sec:5}.

\section{qutrit example}
\label{sec:2}
Before starting to solve the general problem, we first take a look at $2$ simple examples, $2$ and $3$-extension of qutrit.
In fact, these $2$ case clearly demonstrated why our new algorithm can greatly reduce the size of searching space.
We are going to investigate how many real parameters are needed to fully describe the global symmetric extended matrix $\rho_{A\vec{B}}$, which lies in Hilbert space $V=V_A\otimes \mathcal{T}$ constituted by part $A$ and $\vec{B}$, with the constrains that $\Tr_{(AB_1)^c}(\rho_{A\vec{B}})=\rho_{AB}$.

\subsection{$2$-qutrit}
In this case, $\mathcal{T}\equiv V^{(1)}\otimes V^{(2)}$ is constituted by $2$-qutrit $B_1$ and $B_2$, where $V^{(1)}$ and $V^{(2)}$ represents $B_1$ and $B_2$, respectively.
$\mathcal{T}$ is spanned by nine vectors $\{|00\rangle, |01\rangle, \cdots, |22\rangle \}$.
According to different permutation symmetry, $\mathcal{T}$ can be decomposed as $2$ invariant orthogonal subspace, $6$-dimensional bosonic one $V^B$ and $3$-dimensional fermionic one $V^F$.
It is clear that, there does not exist any cross term of bosonic subspace and formionic subspace, therefore we only have to consider the density matrix $\bar{\rho}_{A\vec{B}}$ supporting on $\mathrm{End}(V_A)\otimes \mathrm{End}(V^B)$ and $\tilde{\rho}_{A\vec{B}}$ supporting on $\mathrm{End}(V_A)\otimes \mathrm{End}(V^F)$.

The general form of $\bar{\rho}_{A\vec{B}}$ reads
\begin{equation}\label{Eq: bosonic_general}
  \bar{\rho}_{A\vec{B}}=\sum_{\alpha,\beta=1}^{6}\rho_A^{(\alpha,\beta)}\otimes
  \bar{p}_{\alpha,\beta}|\phi^{\alpha}_{\vec{B}}\rangle\langle\ \phi^{\beta}_{\vec{B}}|,
\end{equation}
where $\rho_A^{(\alpha,\beta)}\in \mathrm{End}(V_A)$, $\bar{p}_{\alpha,\beta}$ is complex number and
\begin{eqnarray*}
  \{|\phi^{\alpha}_{\vec{B}}\rangle\}&=&
  \left\{|00\rangle, |11\rangle, |22\rangle, \frac{1}{\sqrt{2}}(|01\rangle+|10\rangle),\right.\\
  &&\left.\frac{1}{\sqrt{2}}(|02\rangle+|20\rangle), \frac{1}{\sqrt{2}}(|12\rangle+|21\rangle),\right\}.
\end{eqnarray*}
The reduced density matrix can be obtained by doing partial trace over $B_2$
\begin{eqnarray}\label{Eq: bosonic_reduced}
  2\times \Tr_{V^{(2)}}(\bar{\rho}_{A\vec{B}})=\sum_{a,b=0}^2 \bar{M}_{ab}\otimes |a\rangle\langle b|,
\end{eqnarray}
where $\bar{M}_{ab}$ is given by
\begin{eqnarray}
    \notag \bar{M}_{00}&=&2\bar{p}_{1,1}\rho_A^{(1,1)}+\bar{p}_{4,4}\rho_A^{(4,4)}+\bar{p}_{5,5}\rho_A^{(5,5)},\\
    \notag \bar{M}_{11}&=&2\bar{p}_{2,2}\rho_A^{(2,2)}+\bar{p}_{4,4}\rho_A^{(4,4)}+\bar{p}_{6,6}\rho_A^{(6,6)},\\
    \notag \bar{M}_{22}&=&2\bar{p}_{3,3}\rho_A^{(3,3)}+\bar{p}_{5,5}\rho_A^{(5,5)}+\bar{p}_{6,6}\rho_A^{(6,6)},\\
    \notag \bar{M}_{01}&=&{\bar{M}_{10}}^{\dagger}=\sqrt{2}\bar{p}_{4,1}\rho_A^{(4,1)}+\sqrt{2}\bar{p}_{2,4}\rho_A^{(2,4)}+\bar{p}_{6,5}\rho_A^{(6,5)},\\
    \notag \bar{M}_{02}&=&{\bar{M}_{20}}^{\dagger}=\sqrt{2}\bar{p}_{5,1}\rho_A^{(5,1)}+\sqrt{2}\bar{p}_{3,5}\rho_A^{(3,5)}+\bar{p}_{6,4}\rho_A^{(6,4)},\\
    \notag \bar{M}_{12}&=&{\bar{M}_{21}}^{\dagger}=\sqrt{2}\bar{p}_{6,2}\rho_A^{(6,2)}+\sqrt{2}\bar{p}_{3,6}\rho_A^{(3,6)}+\bar{p}_{5,4}\rho_A^{(5,4)}.\\
\end{eqnarray}

It is noticed that in each term of the RHS in Eq(\ref{Eq: bosonic_reduced}) does not contain every $\bar{p}_{\alpha,\beta}$.
In fact, the nonzero coefficients before $\bar{p}_{\alpha,\beta}$ in the term $|a\rangle\langle b|$ are exactly the nonzero entries of representation matrix $T_{(ab)}$ over this bosonic invariant subspace.\footnote{You may find the representation of $su(3)$ for this case and the following case in standard textbook on group theory, such as ~\cite{georgi1983lie, hamermesh2012group}.}

Similarly, one can write down the general form of $\tilde{\rho}_{A\vec{B}}$ supporting on $\mathrm{End}(V_A)\otimes \mathrm{End}(V^F)$
\begin{equation}\label{Eq: fermionic_general}
  \tilde{\rho}_{A\vec{B}}=\sum_{\alpha,\beta=1}^{3}\rho_A^{(\alpha,\beta)}\otimes
  \tilde{p}_{\alpha,\beta}|\psi^{\alpha}_{\vec{B}}\rangle\langle\psi^{\beta}_{\vec{B}}|,
\end{equation}
where $\rho_A^{(\alpha,\beta)}\in \mathrm{End}(V_A)$, $\tilde{p}_{\alpha,\beta}$ is complex number and $\{|\psi^{\alpha}_{\vec{B}}\rangle\}$ is
\begin{eqnarray*}
  \left\{\frac{1}{\sqrt{2}}(|01\rangle-|10\rangle),
  \frac{1}{\sqrt{2}}(|02\rangle-|20\rangle), \frac{1}{\sqrt{2}}(|12\rangle-|21\rangle),\right\}.
\end{eqnarray*}
The reduced density matrix can be obtained by doing partial trace over $B_2$
\begin{eqnarray}\label{Eq: fermionic_reduced}
  2\times \Tr_{V^{(2)}}(\tilde{\rho}_{A\vec{B}})=\sum_{a,b=0}^2 \tilde{M}_{ab}\otimes |a\rangle\langle b|,
\end{eqnarray}
where $\tilde{M}_{ab}$ is given by
\begin{eqnarray}
    \notag \tilde{M}_{00}&=&2\tilde{p}_{1,1}\rho_A^{(1,1)}+\tilde{p}_{2,2}\rho_A^{(2,2)},\\
    \notag \tilde{M}_{11}&=&\tilde{p}_{1,1}\rho_A^{(1,1)}+\tilde{p}_{3,3}\rho_A^{(3,3)},\\
    \notag \tilde{M}_{22}&=&\tilde{p}_{2,2}\rho_A^{(2,2)}+\tilde{p}_{3,3}\rho_A^{(3,3)},\\
    \notag \tilde{M}_{01}&=&{\tilde{M}_{10}}^{\dagger}=\tilde{p}_{3,2}\rho_A^{(3,2)},\\
    \notag \tilde{M}_{02}&=&{\tilde{M}_{20}}^{\dagger}=
    -\tilde{p}_{3,1}\rho_A^{(3,1)},\\
    \notag \tilde{M}_{12}&=&{\tilde{M}_{21}}^{\dagger}=\tilde{p}_{2,1}\rho_A^{(2,1)}.\\
\end{eqnarray}

Taking both $\bar{\rho}_{A\vec{B}}$ and $\tilde{\rho}_{A\vec{B}}$ into account, much less real parameters is needed: the original algorithm searches the entire Hilbert space and thus the number of parameters is $9^2 d_A^2$, while usage of permutation symmetry can reduce such number to $(6^2+3^2)d_A^2$. It should be stressed that such simplification comes from the fact that the cross terms between subspaces corresponding to different permutation symmetry are forbidden.

However, naively usage of symmetry, such as simply symmetrizing the Gell-Mann matrices over $B_1$ and $B_2$ has to determine the positive definiteness of $\textbf{1}$ matrix with dimension $(6^2+3^2)d_A^2$. As a comparison, our method involves determining the positive definiteness of $\textbf{2}$ matrices, whose dimension are $6^2 d_A^2$ and $3^2 d_A^2$ respectively.

\subsection{$3$-qutrit}
In this case, $\mathcal{T}\equiv V^{(1)}\otimes V^{(2)}\otimes V^{(3)}$ is constituted by $3$-qutrit $B_1$, $B_2$ and $B_3$, where $V^{(i)}$ represents $B_i$ respectively.
Similar as the procedure in the previous subsection, one can decompose $\mathcal{T}$ as direct sum of subspace according to different permutation symmetry, and further more, there does not exist cross term between subspaces corresponding to different permutation symmetry.
The permutation symmetry of $3$-qutrit case is much more complicated than the $2$-qutrit case.
It is easy to verify that there exist a $10$-dimensional bosonic subspace and a $1$-dimensional fermionic subspace. Therefore one can solve this problem by imitating previous subsection and obtain the constrain equations.
In this situation, the dimension of searching space can be reduced from $27^2d_A^2$ to $(10^2+16^2+1^2)d^2_A$.

However, more room is left in simplification.
According to Weyl duality, the $16$-dimensional subspace can be further decomposed as $2$ orthogonal $8$-dimensional invariant subspaces $\mathcal{T}^{[2,1]}_1$ and $\mathcal{T}^{[2,1]}_2$, and both are loaded with an equivalent $su(3)$ irreducible representation, described by a two row Young diagram $[2,1]$.\footnote{Here $[\lambda]\equiv \{\lambda_1,\lambda_2,\cdots,\lambda_n\}$ is a partition of integer $k$, where all $\lambda_i$ are integers satisfying $\lambda_1\geq \lambda_2\geq\cdots\geq\lambda_n \geq 0, \sum_{i=1}^n\lambda_i=k$. Such partition is denoted by an $n$-row Young diagram.}

$\mathcal{T}^{[2,1]}_1$ and $\mathcal{T}^{[2,1]}_2$ are spanned by vectors $\{\varphi^{\alpha,(1)}_{\vec{B}}\}$ and $\{\varphi^{\alpha,(2)}_{\vec{B}}\}$:
\begin{widetext}
\begin{eqnarray}\label{Eq: varphi_1}
 \notag &&\left\{\frac{1}{\sqrt{6}}(2|001\rangle-|010\rangle-|100\rangle), \frac{1}{\sqrt{6}}(2|002\rangle-|020\rangle-|200\rangle), \frac{1}{\sqrt{6}}(|011\rangle-|101\rangle-2|110\rangle), \right.\\
 \notag &&\quad \frac{1}{\sqrt{12}}(2|012\rangle-|021\rangle+2|102\rangle
 -|120\rangle-|201\rangle-|210\rangle), \frac{1}{\sqrt{4}}(|021\rangle-|120\rangle+|201\rangle-|210\rangle), \\
 \notag &&\quad \left. \frac{1}{\sqrt{6}}(|022\rangle+|202\rangle-2|220\rangle),
 \frac{1}{\sqrt{6}}(2|112\rangle-|121\rangle-|211\rangle), \frac{1}{\sqrt{6}}(|122\rangle+|212\rangle-2|221\rangle)\right\} ,
\end{eqnarray}

\begin{eqnarray}\label{Eq: varphi_2}
 \notag &&\left\{\frac{1}{\sqrt{2}}(|010\rangle-|100\rangle),
 \frac{1}{\sqrt{2}}(|020\rangle-|200\rangle),
 \frac{1}{\sqrt{2}}(|011\rangle-|101\rangle),\right. \\
 \notag &&\quad \frac{1}{\sqrt{4}}(|021\rangle+|120\rangle-|201\rangle-|210\rangle),
 \frac{1}{\sqrt{12}}(2|012\rangle+|021\rangle-2|102\rangle  -|120\rangle-|201\rangle+|210\rangle), \\
 \notag &&\quad \left.\frac{1}{\sqrt{2}}(|022\rangle-|202\rangle), \frac{1}{\sqrt{2}}(|121\rangle-|211\rangle), \frac{1}{\sqrt{2}}(|122\rangle-|212\rangle),\right\}.
\end{eqnarray}
\end{widetext}
Under the constrain condition imposed in Eq(\ref{Eq: constrain}), the general form of global state $\hat{\rho}_{A\vec{B}}$ supporting on $\mathrm{End}(V_A)\otimes \mathrm{End}(\mathcal{T}^{[2,1]}_1\oplus\mathcal{T}^{[2,1]}_2)$ must be of the following form:
\begin{equation}\label{Eq: mixed_general}
  \hat{\rho}_{A\vec{B}}=\sum_{\alpha,\beta=1}^{8}\rho_A^{(\alpha,\beta)}\otimes
  \hat{p}_{\alpha,\beta}\left(|\varphi^{\alpha,(1)}_{\vec{B}}\rangle\langle\varphi^{\beta,(1)}_{\vec{B}}|
  +|\varphi^{\alpha,(2)}_{\vec{B}}\rangle\langle\varphi^{\beta,(2)}_{\vec{B}}|\right),
\end{equation}
The reduced density matrix can be obtained by doing partial trace over $B_2$
\begin{eqnarray}\label{Eq: mixed_reduced}
  && \frac{3}{2}\times \Tr_{V^{(2)}}(\hat{\rho}_{A\vec{B}})=\sum_{a,b=0}^2 \hat{M}_{ab}\otimes |a\rangle\langle b|,
\end{eqnarray}

where $\tilde{M}_{ab}$ is given by
\begin{widetext}
\begin{eqnarray}
  \notag \hat{M}_{00}&=&2\hat{p}_{1,1}\rho_A^{(1,1)}+2\hat{p}_{2,2}\rho_A^{(2,2)}+\hat{p}_{1,1}\rho_A^{(1,1)}  +\hat{p}_{4,4}\rho_A^{(4,4)}+p_{5,5}\rho_A^{(5,5)}+\hat{p}_{6,6}\rho_A^{(6,6)},\\
  \notag \hat{M}_{11}&=&\hat{p}_{1,1}\rho_A^{(1,1)}+2\hat{p}_{3,3}\rho_A^{(3,3)}+\hat{p}_{4,4}\rho_A^{(4,4)}\hat{p}_{5,5}\rho_A^{(5,5)}+2\hat{p}_{7,7}\rho_A^{(7,7)}+\hat{p}_{8,8}\rho_A^{(8,8)}, \\
  \notag \hat{M}_{22}&=&\hat{p}_{2,2}\rho_A^{(2,2)}+\hat{p}_{4,4}\rho_A^{(4,4)}+\hat{p}_{5,5}\rho_A^{(5,5)}+2\hat{p}_{6,6}\rho_A^{(6,6)}+\hat{p}_{7,7}\rho_A^{(7,7)}+2\hat{p}_{8,8}\rho_A^{(8,8)},   \\
  \notag \hat{M}_{01}=\hat{M}_{10}^{\dagger}&=&-\frac{1}{\sqrt{2}}\hat{p}_{4,1}\rho_A^{(4,1)}+\sqrt{\frac{3}{2}}\hat{p}_{5,1}\rho_A^{(5,1)}
  +\hat{p}_{6,2}\rho_A^{(6,2)}
  +\frac{1}{\sqrt{2}}\hat{p}_{8,4}\rho_A^{(8,4)}-\sqrt{\frac{3}{2}}\hat{p}_{8,5}\rho_A^{(8,5)}
  -\hat{p}_{7,3}\rho_A^{(7,3)},
  \\
  \notag \hat{M}_{02}=\hat{M}_{20}^{\dagger}&=&
  \hat{p}_{2,1}\rho_A^{(2,1)}+\frac{1}{\sqrt{2}}\hat{p}_{4,3}\rho_A^{(4,3)}
  +\sqrt{\frac{3}{2}}\hat{p}_{5,3}\rho_A^{(5,3)}+\frac{1}{\sqrt{2}}\hat{p}_{6,4}\rho_A^{(6,4)}
  +\sqrt{\frac{3}{2}}\hat{p}_{6,5}\rho_A^{(6,5)}+\hat{p}_{2,1}\rho_A^{(2,1)},\\
 \notag
 \hat{M}_{12}=\hat{M}_{21}^{\dagger}&=&\hat{p}_{3,1}\rho_A^{(3,1)}+\sqrt{2}\hat{p}_{4,2}\rho_A^{(4,2)}+\hat{p}_{7,4}\rho_A^{(7,4)}+\hat{p}_{8,7}\rho_A^{(8,7)}.\\
\end{eqnarray}
\end{widetext}

Due to the permutation requirement, the number of real parameters is less than $(10^2+8^2+1^2)d_A^2$, since some pairs of $\alpha$ and $\beta$ may contribute nothing when computing the $1$-body reduced density matrix.\footnote{BUt these number can not be set $0$ directly, since they may affect the positive definiteness.}
It should be stressed that the simplification comes from not only the fact that the cross terms between subspaces corresponding to different permutation symmetry are forbidden, but also arises from the fact that majority of cross terms within subspaces corresponding to identical permutation symmetry are also forbidden.
It should also be noticed that, via our method, one can check the positive definiteness of global state by successively checking the positive definiteness of density matrix corresponding to different permutation symmetries.

\section{complexity of improved SDP}
\label{sec:3}
In this section we are going to give the general form of a global state that corresponds to the given quantum marginals $\rho_{AB}$.

Consider the symmetric extension problem described in Eq.(\ref{Eq: SDP_SE}).
It is required that the global state $\rho_{AB_1\cdots B_k}$ is invariant under any exchange of $B_i$ and $B_j$, but it does not require that $\rho_{AB_1\cdots B_k}$ must support on a subspace with specific permutation symmetry.
e.g. for a $2$-symmetric extendible state, its extension can be bosonic, which supports on the symmetric subspace only, or fermionic, whose support only resides on the antisymmetric subspace, or more generally, can be a mixture of both.

Consider a Hilbert space $\ct=\bigotimes_{i=1}^k V^{(i)}$ constituted by $k$-qudit $B_1, B_2,\cdots, B_k$, whose computational basis is $\left\{\Phi_{i_1,i_2,\cdots,i_k}\equiv\ket{i_1,i_2,\cdots,i_k}\right\}$, where $i_1,i_2,\cdots,i_k=0,1,\cdots,d-1$.

Each subsystem $V^{(i)}$ is invariant under $SU(d)$ 'rotation', and transforms according to the $d$ dimensional fundamental irreducible representation $D^{[1]}$, which corresponds to the Young diagram $[1]$. \footnote{this is the $1$ block Young diagram \Yvcentermath1
\young({\,}).}

Therefore the Lie algebra $su(d)$, which is constituted by $3$ series of zero trace hermitian matrices and describes the infinite small rotation of $SU(d)$, has the following matrix forms on each identical $V^{(j)}$, if we set $\ket{i}$ to be the natural basis,
\begin{eqnarray*}
\left(T^{(1)}_{mn}\right)_{st} &=& \frac{1}{2}(\delta_{ms}\delta_{nt}+\delta_{ns}\delta_{mt}),\\
\left(T^{(2)}_{mn}\right)_{st} &=& \frac{-i}{2}(\delta_{ms}\delta_{nt}+\delta_{ns}\delta_{mt}),\\
\left(T^{(3)}_{p}\right)_{st} &=& \left\{
\begin{tabular}{l l}
$\delta_{st}[2(p+1)p]^{-\frac{1}{2}}$, & $s < p$,\\
$-\delta_{st}[p/(2p+2)]^{\frac{1}{2}}$, & $s = p$,\\
$0$, & $s > p$,
\end{tabular}
\right.,
\end{eqnarray*}
where $m<n,$ and $1\leq p\leq d-1$.
Taking the global phase into account, one should also include the identity matrix.
Therefore one can obtain a new basis for Lie algebra $u(d)$ by
\[\{T_{ab}|\left(T_{ab}\right)_{st}=\delta_{as}\delta_{bt}, 0\leq a,b\leq d-1\}.\]

$\ct$ is also invariant under the global $U(d)$ transformation , whose corresponding Lie algebra is given by $\{\mathbf{T}_{ab}|\mathbf{T}_{ab}=\sum_i T^{(i)}_{ab}\}$.\footnote{Here $T^{(i)}_{ab}$ denotes that the $i$-th subsystem transforms according to $T_{ab}$ while others according to the identity operator.}
$\ct$ transforms under representation $\otimes^k[1]$, which is not irreducible, but can be decomposed as direct sum of a series of irreducible representations,
\begin{eqnarray}\label{multiplicity}
\bigotimes\nolimits^k D^{[1]}=\bigoplus_{[\lambda]}m_{[\lambda]} D^{[\lambda]},
\end{eqnarray}
where $m_{[\lambda]}$ is the multiplicity of irreducible representation $D^{[\lambda]}$.
This is equivalent to say that $\ct$ can be partitioned as direct sum of subspaces.\footnote{Please be notified that, subspaces corresponding to different Young diagram are orthogonal to each other, while those corresponding to same Young diagram are not. However, it is guaranteed that the intersection of such $2$ different subspaces is zero.}
\begin{equation}
\ct=\bigoplus_{[\lambda]} m_{[\lambda]} \ct^{[\lambda]}.
\end{equation}

It can be easily manifested that, such $\ct^{[\lambda]}$ has particular permutation symmetry described by Young diagram $[\lambda]$, and the multiplicity $m_{[\lambda]}$ equals to the dimension of irreducible representation of $S_k$ corresponding to the identical Young diagram $[\lambda]$, which gives an equation
\begin{equation}\label{dimension relation}
  d^k=\sum_{[\lambda]}m_{[\lambda]}D^{[\lambda]}.
\end{equation}

Two irreducible representation spaces $\ct^{[\lambda]}_{\mu}$ and $\ct^{[\lambda]}_{\nu}$ corresponding to same Young diagram but different Young tableaus are orthogonal to each other. Although there might probably be multiplicity in some weight subspace for a general irreducible subspace, one can uniquely label a vector within arbitrary given irreducible subspace by its weight $\vec{\omega}$ in $su(d)$ and the subgroup chain $su(d)\supset su(d-1)\supset \cdots\supset su(2)$~\cite{gelfand1988collected}.
Thus one can safely use weight $\vec{\omega}$ to label different states inside an irreducible subspace $\ct^{[\lambda]}_{\mu}$.
Therefore, $\{\ket{[\lambda],\mu,\vec{\omega}}\}$ labels a complete basis of $\ct$ one by one, where $[\lambda]$ tells inequivalent $su(d)$ representations while $\mu$ differentiate equivalent ones.
They together determine an orthogonal irreducible subspace, and $\vec{\omega}$ labels every different vectors inside.

On the other hand, $\{\ket{[\lambda],\mu,\omega}\}$ can be interpreted in another way:
$\vec{\omega}$ describes the weight, $[\lambda]$ tells inequivalent $S_k$ representations, thus these two parameter differentiate orthogonal invariant subspaces, while $\mu$ labels vectors inside.
From now on we shall use $\ket{\vec{\omega}^{[\lambda]}_{\mu}}$ short for $\ket{\vec{\omega},[\lambda],\mu}$.

Any matrix $\rho_{A B_1 B_2 \cdots B_n} \in \mathrm{End}(V_A) \otimes \mathrm{End}(\ct)$ can be expressed as
\begin{eqnarray}\label{general_form}
\rho_{A B_1 B_2 \cdots B_k}&=&\sum_{\alpha,\alpha'}\sum_{[\lambda],[\lambda']}\sum_{\mu,\mu'}\sum_{\vec{\omega},\vec{\omega'}}
\ket{\psi^{\alpha}_{\vec{\omega},[\lambda],\mu}}\bra{\psi^{\alpha'}_{\vec{\omega'},[\lambda'],\mu'}}\notag\\
&&\qquad \otimes\ket{\vec{\omega}^{[\lambda]}_{\mu}}\bra{\vec{\omega'}^{[\lambda']}_{\mu'}},\notag\\
\end{eqnarray}
where $\ket{\psi^{\alpha}_{\vec{\omega},[\lambda],\mu}}$ are non-normalized state in $V_A$ and $\alpha$ label different states in $V_A$.

Insert Eq.(\ref{general_form}) into Eq.(\ref{Eq: SDP_SE}). $\forall \pi \in S_k$ we get a series of constrains for $\rho_{A B_1 B_2 \cdots B_k}$:
\begin{eqnarray}
&&\quad\forall [\lambda],[\lambda']\vec{\omega},\vec{\omega'}\textrm{ and }\mu,\mu',\notag\\
&&\sum_{\alpha,\alpha'}\ket{\psi^{\alpha}_{\vec{\omega},[\lambda],\mu}}\bra{\psi^{\alpha'}_{\vec{\omega'},[\lambda'],\mu'}}
\sum_{\nu,\nu'}
\ca(\pi)^{[\lambda]}_{\mu,\nu}
\ca(\pi)^{[\lambda']*}_{\nu',\mu'}
\ket{\vec{\omega}^{[\lambda]}_{\nu}}
\bra{\vec{\omega'}^{[\lambda']}_{\nu'}}\notag\\
&&\qquad=
\sum_{\alpha,\alpha'}\ket{\psi^{\alpha}_{\vec{\omega},[\lambda],\mu}}\bra{\psi^{\alpha'}_{\vec{\omega'},[\lambda'],\mu'}}
\ket{\vec{\omega}^{[\lambda]}_{\mu}}\bra{\vec{\omega'}^{[\lambda']}_{\mu'}},\qquad
\end{eqnarray}
where $\ca^{[\lambda]}$ and $\ca^{[\lambda']}$ are irreducible representations of permutation group $S_k$.

Define matrix
\begin{eqnarray}
&&M(\vec{\omega},\vec{\omega'},[\lambda],[\lambda'])\nonumber\\
&\equiv&
\sum_{\mu,\mu'}M(\vec{\omega},\vec{\omega'},[\lambda],[\lambda'])_{\mu\mu'}
\ket{\vec{\omega}^{[\lambda]}_{\mu}}\bra{\vec{\omega'}^{[\lambda']}_{\mu'}},
\end{eqnarray}
where
\begin{equation}
M(\vec{\omega},\vec{\omega'},[\lambda],[\lambda'])_{\mu\mu'} \equiv\sum_{\alpha,\alpha'} \ket{\psi^{\alpha}_{\vec{\omega},[\lambda],\mu}}\bra{\psi^{\alpha'}_{\vec{\omega'},[\lambda'],\mu'}},
\end{equation}
thus $\forall \pi \in S_k$
\begin{eqnarray}
\label{commute}
&&\ca^{[\lambda]}(\pi)M(\vec{\omega},\vec{\omega'},[\lambda],[\lambda'])\ca^{[\lambda']}(\pi)^{\dagger}\nonumber\\
&=&M(\vec{\omega},\vec{\omega'},[\lambda],[\lambda']).\notag\\
\end{eqnarray}

Schur's lemma guarantee that,
\begin{itemize}
\item[a.] when $[\lambda] \neq [\lambda']$, $M=0$;
\item[b.] when $[\lambda] = [\lambda']$, $M$ is invertible.
\end{itemize}

Choose $\ket{\vec{\omega}_{\mu}^{[\lambda]}}$ carefully such that the representation $\ca^{[\lambda]}$ are identical, not just an isomorphic matrix, for different weight $\omega$.
Then all $M(\vec{\omega},\vec{\omega'},[\lambda],[\lambda])$ can be proportional to the corresponding identity matrix.
Therefore, one could eliminate majority of cross terms and restrict $\rho_{A B_1 B_2 \cdots B_k}$ to
\begin{eqnarray}\label{specific form}
\rho_{A B_1 B_2 \cdots B_k}&=&\sum_{[\lambda]}\sum_{\vec{\omega},\vec{\omega'}}
f([\lambda],\vec{\omega},\vec{\omega'})
\sigma([\lambda],\vec{\omega},\vec{\omega'})\notag\\
&&\otimes \sum_{\mu}\ket{\vec{\omega}^{[\lambda]}_{\mu}}\bra{\vec{\omega}'^{[\lambda]}_{\mu}},
\end{eqnarray}
where $f([\lambda],\vec{\omega},\vec{\omega'})$ is the coefficient and $\sigma([\lambda],\vec{\omega},\vec{\omega'})\in \mathrm{End}(V_A)$(do not have to be a density matrix!), both of which correspond to $S_k$ irreducible representation described by Young diagram $[\lambda]$ and different weight $\vec{\omega}$ and $\vec{\omega'}$.

Our next task is to determine the RDM of global state given by Eq.(\ref{specific form}).
For every given $[\lambda],\vec{\omega}$ and $\vec{\omega'}$, one could temporally ignore system $A$ and concentrate on group $\{B_1,B_2,\cdots,B_k\}$,
\begin{eqnarray}\label{partial trace}
  \notag
  \sum_{i,j=0}^{d-1}B_{ij}\ket{i}\bra{j}
  &=&
  \Tr_{B_1^c}\left(\sum_{\mu}\ket{\vec{\omega}^{[\lambda]}_{\mu}}\bra{\vec{\omega}'^{[\lambda]}_{\mu}}\right)\\
  &=&
   \notag \sum_{i,j=0}^{d-1}\ket{i}\bra{j}\Tr\left(T^{(1)}_{ji}\sum_{\mu}\ket{\vec{\omega}^{[\lambda]}_{\mu}}\bra{\vec{\omega}'^{[\lambda]}_{\mu}}\right)\\
   &=&
   \notag
   \sum_{i,j=0}^{d-1}\ket{i}\bra{j}
   \Tr\left(\frac{1}{k}\mathbf{T}_{ji}\sum_{\mu}
   \ket{\vec{\omega}^{[\lambda]}_{\mu}}\bra{\vec{\omega}'^{[\lambda]}_{\mu}}\right)\\
   \notag &=&
   \sum_{i,j=0}^{d-1}\ket{i}\bra{j}
   \left(\frac{m_{[\lambda]}}{k}\bra{\vec{\omega}'^{[\lambda]}}\mathbf{T}_{ji}\ket{\vec{\omega}^{[\lambda]}}\right),\\
\end{eqnarray}
where $\bra{\vec{\omega}'^{[\lambda]}}\mathbf{T}_{ji}\ket{\vec{\omega}^{[\lambda]}}$ is exactly the matrix element of irreducible representation corresponding to Young diagram $[\lambda]$ for generator $T_{ji}$ in Lie algebra $u(d)$.\footnote{Do not worry about this part, the general matrix form of $\mathbf{T}_{ji}$ in irreducible representation $D^{[\lambda]}$ of $u(d)$ have been calculated by mathematicians and you can refer to~\cite{gelfand1988collected}.}
Taking subsystem $A$ into account, one can obtain
\begin{eqnarray}\label{final}
  \notag&&\Tr_{(AB_1)^c}\left(\rho_{A B_1 B_2 \cdots B_k}\right)\\
  &=& \notag
  \sum_{[\lambda]}\sum_{m,n=0}^{d_A-1}\sum_{i,j=0}^{d-1}\ket{m}\bra{n}\otimes\ket{i}\bra{j}\\
  &&\notag\times\sum_{\vec{\omega},\vec{\omega}'}
  \sigma([\lambda],\vec{\omega},\vec{\omega'})_{mn}
  \frac{m_{[\lambda]}}{k}\bra{\vec{\omega}'^{[\lambda]}}\mathbf{T}_{ji}\ket{\vec{\omega}^{[\lambda]}}
  f([\lambda],\vec{\omega},\vec{\omega'}).
  \\
\end{eqnarray}

For every given $[\lambda]$, the number of different values for $\vec{\omega}$ and $\vec{\omega'}$ is just the dimension of $u(d)$ irreducible representation $D^{[\lambda]}$.
Therefore, ignoring the size of subsystem $A$, the size of searching space in dealing with symmetric extension is given by
\begin{equation}\label{exact size}
\sum\left(D^{[\lambda]}\right)^2=\binom{d^2-1+k}{k}<d^{2k},
\end{equation}
where the summation runs over all possible proper Young diagrams.
One may conclude that, the dimension of entire searching space grows not faster than $O(k^{d^2})$, which is significantly smaller than the original $O(d^{2k})$, therefore the efficiency of SDP could be greatly improved when dealing with symmetric extension problems.

To guarantee the positive definiteness of solved global matrix, one can test whether the density matrix corresponding to different permutation is positive definite respectively, and hence each testing need much less resource.

To investigate the amount of the resource needed for each single testing, one should focus on the growth rate of $D^{[\lambda]}$ and $m_{[\lambda]}$.
The asymptotic behavior of the upper bound of $D^{[\lambda]}$ is given by
\begin{equation}\label{dimension of irr}
  \prod_{m=1}^{d-1}\frac{1}{(m)!}\left(1+\frac{2k}{d(d+1)}\right)^{\frac{d(d-1)}{2}},
\end{equation}
which corresponds to the irreducible representation that satisfies $\lambda_i-\lambda_{i+1}\approx 2k/d(d-1)$~\cite{fulton1997young}.
For a given $k$, the number of different valid Young diagrams whose row is less than or equal to $d$ is hard to compute analytically, but for sufficient large $k$, the asymptotic value is $\frac{1}{d!}\binom{k+d-1}{k}$.
\footnote{To find a partition(not necessarily a partition corresponds to valid Young diagram!) that satisfies $\sum_i^k\lambda_i=k$ is equivalent to insert $d-1$ separators between a line of $k$ balls, which reads $\binom{k+d-1}{k}$. Since some $\lambda_i$s might be identical, the number of valid different Young diagrams should be less than $\frac{1}{d!}\binom{k+d-1}{k}$, but when $k$ is sufficient large, such odd approaches to $0$, so the asymptotic number of valid Young diagrams is given by $\frac{1}{d!}\binom{k+d-1}{k}$.}

\section{Numerical results}\label{sec:4}

First, we apply our algorithm on the famous bipartite Werner state $\rho_{W,d}\left(\alpha\right)\in\mathcal{H}_d\otimes\mathcal{H}_d$
\[ \rho_{W,d}\left(\alpha\right)=\frac{1}{d^{2}-d\alpha}I-\frac{\alpha}{d^{2}-d\alpha}\sum_{ij}\left|ij\right\rangle \left\langle ji\right|,\;\alpha\in\left[-1,1\right]. \]
Previous work \cite{Johnson_2013} proved that the Werner state is $(1,k)$-extendible for $\alpha\in[-1,\frac{k+d^{2}-d}{kd+d-1}]$. As $k$ goes to infinity, it gives the separable Werner state $\alpha\in\left[-1,\frac{1}{d}\right]$. To obtain such a $(1,k)$-extendible boundary $\alpha_k^*$, one can solve the following semi-definite programming
\begin{align*}
    &\max\;c,\\
    s.t.&\begin{cases}
\rho_{AB_1\cdots B_k}\succeq0,\\
\left(\mathbb{1}^A\otimes P_{ij}\right)\rho_{AB_1\cdots B_k}\left(\mathbb{1}^A\otimes P_{ij}\right)=\rho_{AB_1\cdots B_k},\\
\Tr_{B_1^c}\left(\rho_{AB_1\cdots B_k}\right)=\left(1-c\right)\rho_{o}+c\rho_{W,d}(1),
\end{cases}
\end{align*}
with $\rho_0$ denotes the maximally mixed state\footnote{As semi-definite programming requires linear or affine equation constraint, we convert the non-linear expression $\alpha$ in Werner state into a linear interpolation $(1-c)\rho_0+c\rho_{W,d}(1)$ used in optimization.}, and The boundary can be calculated from the optimal value $\alpha_k^*=\frac{c^{*}d}{c^{*}+d-1}$.

\begin{table}[h!]
    \centering
    \begin{tabular}{cccc}
        \hline
        $(d,k)$& QETLAB (s) & irrep (s) & $\alpha_k^*$ \\
        \hline
        $(2,8)$ & 0.19 & 0.16 & 0.588\\
        $(2,10)$ & 12.60 & 0.16 & 0.571\\
        $(2,16)$ & - & 0.32 & 0.545\\
        $(2,32)$ & - & 3.18 & 0.523\\
        $(2,32)$ & - & 51.96 & 0.512\\
        \hline
        $(3,3)$ & 0.62 & 0.51 & 0.818\\
        $(3,4)$ & 7.96 & 2.38 & 0.714\\
        $(3,5)$ & - & 11.56 & 0.647\\
        $(3,6)$ & - & 55.60 & 0.6\\
        \hline
    \end{tabular}
    \caption{\label{table:werner-boundary} Time usage for calculating the Werner $(1,k)$-extendible boundary. The dashed line ``-'' indicates the optimization failed due to memory limitation or intolerable time usage. }
\end{table}

The results are shown in TABLE~\ref{table:werner-boundary}. We compare the time required with the software QETLAB \cite{nathaniel_johnston_2015_14186}, a widely-used MATLAB package in quantum information community. The benchmark is performed on a standard laptop, AMD R7-5800H, 16 CPU cores (hyperthread enabled), 16GB memory and our algorithm is implemented in CVXPY package \cite{diamond2016cvxpy} with MOSEK solver \cite{mosek}. The solved boundary $\alpha_k^*$ is within $10^{-8}$ absolute error compared with the analytical results. From the results, A significant speedup can be observed and a much larger $k$-extension problem can be handled for our algorithm.

We explicitly calculate the dimension of searching space and the number of parameters required to be tested for positive definiteness, which demonstrate the efficiency of our algorithms, as shown in TABLE~\ref{table:number-of-parameter}. \footnote{Our algorithm needs to undergo multiple different positive-definiteness tests, where each different Young diagram corresponds to its own test, but each individual test involves significantly small matrix, and hence the efficiency will be improved.}

\begin{table*}[ht!]
\centering
\begin{tabular}{ccc}
    \hline
    $(d,k)$  & \#searching space (QETLAB,irrep) & \#positive definiteness (QETLAB,irrep)\\
    \hline
    $(3,3)$ & $(729,165)$  & $(729,10^2+8^2+1^2)$ \\
    $(3,4)$ & $(6561,495)$ & $(6561,15^2+15^2+6^2+3^2)$ \\
    $(3,5)$ & $(59049,1287)$ & $(59049,21^2+24^2+15^2+6^2+3^2)$  \\
    $(3,6)$ & $(531441,3003)$ & $(531441,28^2+35^2+27^2+10^2+10^2+8^2+1^2)$ \\
    $(4,3)$ & $(4096,816)$ & $(4096,20^2+20^2+4^2)$ \\
    $(4,4)$ & $(65536,3876)$ & $(65536,35^2+45^2+20^2+15^2+1^2)$ \\
    $(4,5)$ & $(1048576,15504)$ & $(1048576,56^2+84^2+60^2+36^2+20^2+4^2)$ \\
    \hline
\end{tabular}
\caption{\label{table:number-of-parameter} Dimensions and number of parameters needed in positive definiteness.}
\end{table*}

\section{discussion}
\label{sec:5}
The complexity of our new algorithm for dealing with k-symmetric extensions of quantum states is $O(k^{d^2})$, which is an improvement over the original algorithm with $O(d^{2k})$ complexity.
However, it is important to note that the complexity of detecting entanglement is a QMA problem, which means that it is generally considered to be computationally hard.
Although our new algorithm reduces the computational complexity of the problem, it does \emph{not} change the fundamental difficulty of detecting entanglement.
This is due to the fact that the size of the input of this problem is given by $O(\log k, d)$, and hence the resources needed in our algorithm still grows exponentially relative to the input.
Therefore, while our algorithm presents some advance, it does not contradict the known fact that detecting entanglement is a QMA problem.
The challenge of detecting entanglement remains an important area of research, with many open questions and opportunities for new breakthroughs.

\section*{Acknowledgments}
The authors would like to thank for Ruan Dong, Huang Huajun, Huang Shilin for help discussion.
Y-N.Li is supported by National Natural Science Foundation of China under Grant No. 12005295
S-Y.Hou is supported by National Natural Science Foundation of China under Grant No. 12105195. C.Zhang, Z-P.Wu, and B.Zeng are supported by GRF (grant no. 16305121).

\bibliographystyle{apsrev4-1}
\bibliography{SDP_for_Symmetric_Extension}

\end{document}